\documentclass[apj,twocolumn]{emulateapj}

\usepackage{amsmath}

\slugcomment{Accepted for publication in ApJ}

\shortauthors{G. Parmentier \& A. Pasquali}
\shorttitle{A new star formation rate-dense gas mass relation}

\newcommand\effint{\epsilon_{\rm ff, int}}
\newcommand\effmeas{\epsilon_{\rm ff, meas}}

\newcommand\tff{\tau_{\rm ff}}

\newcommand\Ms{M_{\odot}}

\newcommand\Mspp{M_{\odot} \cdot pc^{-2}}
\newcommand\Msppp{M_{\odot} \cdot pc^{-3}}

\newcommand\cc{cm^{-3}}

\newcommand\fft{freefall time }
\newcommand\sfe{star formation efficiency }
\newcommand\sfr{star formation rate }
\newcommand\stf{star formation }
\newcommand\sfing {star-forming }

\newcommand\Son{Solar neighborhood }

\begin{document}



\title{A new parameterization of the star formation rate-dense gas mass relation: \\ embracing gas density gradients}


\author{G.~Parmentier\altaffilmark{1} \& A.~Pasquali\altaffilmark{1}}


\altaffiltext{1}{Astronomisches Rechen-Institut, Zentrum f\"ur Astronomie der Universit\"at Heidelberg, M\"onchhofstr. 12-14, D-69120 Heidelberg, Germany}


\begin{abstract}
It is well-established that a gas density gradient inside molecular clouds and clumps raises their \sfr compared to what they would experience from a gas reservoir of uniform density.  This effect should be observed in the relation between dense-gas mass $M_{dg}$ and \sfr $SFR$ of molecular clouds and clumps, with steeper gas density gradients yielding higher $SFR/M_{dg}$ ratios.  The content of this paper is two-fold.  Firstly, we build on the notion of magnification factor introduced by \citet{par19} to redefine the dense-gas relation (i.e. the relation between $M_{dg}$ and $SFR$).  Not only does the $SFR/M_{dg}$ ratio depend on the mean \fft of the gas and on its (intrinsic) \sfe per free-fall time, it also depends on the logarithmic slope $-p$ of the gas density profile and on the relative extent of the constant-density region at the clump center.  Secondly, we show that nearby molecular clouds follow the newly-defined dense-gas relation, provided that their dense-gas mass is defined based on a volume density criterion.  We also find the same trend for the dense molecular clouds of the Central Molecular Zone (CMZ) of the Galaxy, although this one is scaled down by a factor of $10$ compared to nearby clouds.  The respective locii of both nearby and CMZ clouds in the $(p, SFR/M_{dg})$ parameter space is discussed.  
\end{abstract}



\section{Introduction} \label{sec:intro}

Local galaxies, molecular clumps of the Galactic disk and molecular clouds of the Solar neighborhood present a linear correlation between their dense gas mass, $M_{dg}$, and \stf rate, $SFR$ \citep{gs04,wu10,lad10}. 
This may suggest that the activity of a \sfing region is controlled by its  dense-gas content, usually defined as gas with a number density higher than $\sim 10^4\,\cc$, equivalent to a volume density higher than $\sim 700\,\Msppp$.  
The dense gas relation, that is, $SFR$ in dependence of $M_{dg}$, comes with a scatter, however, and this scatter does not result from random observational uncertainties only.  \citet{par19} builds on the data of \citet{kai14} for nearby molecular clouds to show that clouds with steeper gas density gradients tend to experience higher \stf rates per unit mass of their dense gas (see her Fig.~1).  This is in line with the conclusions of analytical and semi-analytical calculations \citep{tan06,elm11,par14p,par19}, and of hydrodynamical simulations \citep{cho11,gir11a}, which all demonstrate that a gas density gradient inside molecular clouds and clumps enhance their \stf rate.  This immediately suggests that part of the scatter observed in the dense-gas relation has a physical origin, namely, the variations in the gas spatial distribution from one \sfing region to another.    

\citet{par19} introduces the notion of magnification factor, defined as the ratio between the \sfr of a centrally-concentrated clump, and the \sfr that the clump would experience if its gas was uniformly distributed.  In other words, the magnification factor compares the \sfr of a centrally-concentrated clump with that of its top-hat equivalent, the latter containing inside the same clump radius the same gas mass but distributed following a top-hat volume density profile.  Very steep gas density profiles have been revealed in the dense-gas component of two Galactic molecular clouds \citep{schn15}, with NGC~6334 boasting for its dense gas a power-law volume density profile $\rho_{gas}(r) \propto  r^{-p}$ as steep as $p=4$ ($r$ is the distance to the center of the dense-gas clump and spherical symmetry is assumed).  \citet{par19} shows that such a steep density gradient holds the potential to boost the \sfr of spherical clumps by up to three orders of magnitude compared to their top-hat equivalent.  How high the magnification factor exactly is depends on the shape of the gas density profile at the clump center \citep[see Section~5 in][]{par19}.  In a follow-up study, \citet{par20} details how the relation between the surface densities in gas and in stars measured locally inside \sfing regions can help unlock their magnification factor.  In other words, resolved observations of \sfing regions can reveal how much their initial gas density gradient has contributed to their \stf history.

In this paper, we compare the predictions of our model to the data available for two sets of Galactic molecular clouds.  Firstly, we consider, more deeply than previously done in \citet{par19}, the data of \citet{kai14} for nearby clouds.  Secondly, we consider the data of \citet{lu19} for the dense molecular clouds of the Central Molecular Zone (CMZ).  For these two data sets, $p$ is observed in the range $1 < p < 2.2$.

The outline of the paper is as follow.  In Section~\ref{sec:mot}, we remind the notions of intrinsic \sfe per free-fall time, $\effint$, magnification factor, $\zeta$, and the relation between magnification factor $\zeta$ and gas density profile.  In Section~\ref{sec:dgrel}, we show that the dense-gas mass-SFR relation is more appropriately defined as a permitted area of the $(p,SFR/M_{dg})$ parameter space, in which most of the data of \citet{kai14} lie.  Section~\ref{sec:kai} discusses data and model caveats.  Section~\ref{sec:cmz} compares our model predictions to the data from \citet{lu19} and \citet{kau17b} for the CMZ molecular clouds, and discusses differences and similarities between nearby and CMZ clouds.  A summary and the conclusions are presented in Section~\ref{sec:conc}.

\section{Magnification factor $\zeta$ and gas density gradient}\label{sec:mot}
In this section, we recall how the \sfr and magnification factor of a centrally-concentrated clump are related, as well as the analytical expression of the magnification factor for a pure power-law density profile.  The equations presented here are valid regardless of the clump mean density.  In the next section, we will apply them to the particular case of dense-gas clumps.

The \sfr of a gaseous clump with a density gradient is higher than if that same clump was made of uniform-density gas \citep{tan06,gir11a,elm11,par14p,par19}.  Centrally-concentrated clumps actually process their gas into stars at a pace faster than expected based on their mean free-fall time, because the density of their central regions is higher than the clump gas mean density.  \citet{par19} refers to the ratio between the \sfr of a clump with a density gradient, $SFR_{clump}$, and the \sfr of its top-hat equivalent, $SFR_{TH}$, as the {\it magnification factor} $\zeta$ \citep[see Equation~8 in ][]{par19}:  
\begin{equation}
\zeta = \frac{SFR_{clump}}{SFR_{TH}}\,.
\label{eq:zetaratio}
\end{equation}
The \sfr of a clump can thus be written as \citep[see Section~2 in][for a discussion]{par19}
\begin{equation}
SFR_{clump}=\zeta \effint \frac{m_{gas}}{\langle \tff \rangle}\;.
\label{eq:sfrclump2}
\end{equation}
In this equation, $m_{gas}$ is the gas mass hosted by the clump\footnote{The clump total mass $m_{clump}$ consists of the gas mass $m_{gas}$ and the mass $m_{stars}$ of the stars it has formed.}, and $\langle \tff \rangle$ is the mean \fft of the clump gas:  
\begin{equation}
\langle \tff \rangle = \sqrt{\frac{3\pi}{32G \langle \rho_{gas} \rangle}}\;.
\label{eq:tff}
\end{equation}  
The latter is defined based on the clump gas mean density
\begin{equation}
\langle \rho_{gas} \rangle = \frac{m_{gas}}{\frac{4}{3}\pi r_{clump}^3}\,,
\end{equation}
with $r_{clump}$ the radius containing the clump gas mass $m_{gas}$.
$\zeta$ is the magnification factor, which quantifies by how much the density gradient of a clump enhances its \sfr compared to what is expected  based on its mean free-fall time $\langle \tff \rangle$. 
Equation \ref{eq:sfrclump2} therefore allows one to disentangle the contribution of the density gradient, embodied by the $\zeta$ factor, from that of the \sfe per free-fall time itself.  \citet{par19} coins $\effint$ the {\it intrinsic} \sfe per free-fall time since it is independent of the clump density gradient\footnote{This is as opposed to the (globally-)measured \sfe per freefall time, defined as $\effmeas = \zeta \effint$ \citep{par19}}.  For a top-hat profile, $\zeta=1$ and the \sfr obeys:
\begin{equation}
SFR_{TH}=\effint \frac{m_{gas}}{\langle \tff \rangle}\;.
\label{eq:sfrth}
\end{equation}
Note that the gas density of the top-hat equivalent is equal to the gas mean density $\langle \rho_{gas} \rangle$ of the centrally-concentrated clump (same gas mass $m_{gas}$ enclosed within the same clump radius $r_{clump}$).

\begin{figure}
\begin{center}                 
\epsscale{1.0}  \plotone{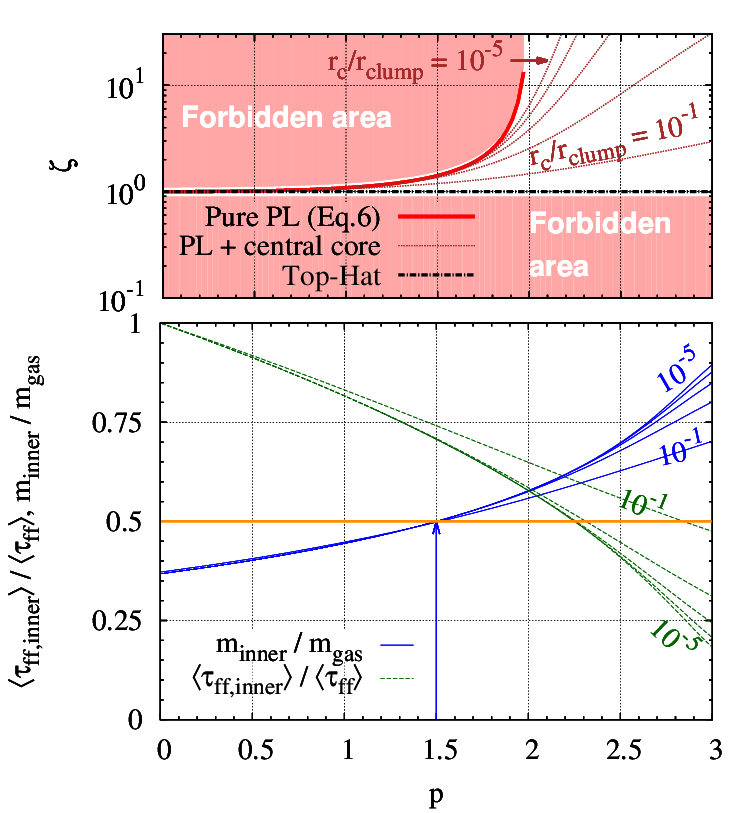}
\caption{
Top panel: The magnification factor $\zeta$ as a function of the steepness $p$ of a power-law density profile.   The cases of a pure power-law (Equation~\ref{eq:zeta}) and of power-laws with a central core are shown.  The core spatial extent is expressed as a fraction of the clump radius (from top to bottom: $r_c/r_{clump} = 10^{-5}$, $10^{-4}$, $10^{-3}$, $10^{-2}$, and $10^{-1}$).  The black dashed line highlights $\zeta =1$, i.e. the case of a clump with a top-hat profile.  The corresponding forbidden regions (see text) are marked as shaded areas.  Bottom panel: mass fraction $m_{inner}/m_{gas}$ of the clump inner region (i.e. the region where the gas has  a density higher than the mean density of the clump gas; blue lines), and ratio between the mean free-fall time of the inner region and the mean free-fall time of the clump gas (green lines).  They are both shown in dependence of $p$ and for the same $r_c/r_{clump}$ ratios as in the top panel.
} 
\label{fig:denser}
\end{center} 
\end{figure}

For a pure power-law gas density profile of steepness $p$, $\rho_{gas}(r) \propto r^{-p}$, $\zeta$ can be obtained analytically when $0\leq p<2$.  $\zeta$ obeys Equation~6 in \citet{par19}, which we give here for the sake of clarity: 
\begin{equation}
\zeta = \frac{(3-p)^{3/2}}{2.6(2-p)}\;.
\label{eq:zeta}
\end{equation}
A similar equation is given by \citet[][their equation 2]{tan06}.  Equation~\ref{eq:zeta} is shown as the solid red line in the top panel of Figure~\ref{fig:denser}.

When $p \geq 2$, the \sfr at the clump center tends towards infinity if the gas density profile is a pure power law (see Appendix).  A central core thus needs to be added to the clump density profile to remove the density singularity and  the magnification factor can then be easily computed by a numerical integration \citep[Eqs~4 and 11 in][]{par19}.  The results are shown as the thin dotted brown lines in the top panel of Fig.~\ref{fig:denser}.  Note that since a central core can be added to any power-law profile, we show the results for $p \geq 0$, and not just for $p\geq 2$.  Each line corresponds to a ratio $r_c/r_{clump}$ of the core radius to the clump radius.  For a given steepness $p$, the larger the ratio $r_{c}/r_{clump}$, the smaller the magnification factor $\zeta$.  This is so because a larger ratio $r_{c}/r_{clump}$ reduces the density contrast between the center and the edge of the clump, resulting thereby in a smaller magnification factor $\zeta$ (Fig.~5 in \citet{par19}, \citet{elm11}).  Note that a relative core size $r_c/r_{clump}>0.1$ damps the magnification factor significantly.

Equation~\ref{eq:zeta}, which has been obtained under the assumption of a pure power-law density profile, therefore defines an upper limit to the magnification factor when $p<2$.  A pure power-law actually maximises the gas density contrast between clump edge and clump center.  As for a lower limit, it is naturally given by $\zeta = 1$, which corresponds to the case of a top-hat profile (or, equivalently, a "central" core covering the full molecular clump, i.e. $r_c >> r_{clump}$).  These two limits (i.e. $\zeta$ as given by Eq.~\ref{eq:zeta} and $\zeta=1$) defines forbidden areas in the $(p,\zeta)$ parameter space: $\zeta$ cannot be higher than given by a pure power-law density profile, and it cannot be lower than found for a top-hat profile.  These forbidden areas are highlighted in light red in the top panel of Fig.~\ref{fig:denser}.  

For a given $r_c/r_{clump}$ ratio, the increase of $\zeta$ with $p$ stems from steeper density gradients having higher mass fractions of gas located in the clump inner regions, where the local \fft is shorter than the mean \fft $\langle \tau_{ff} \rangle$ of the clump gas.
To illustrate this, we now calculate the mass fraction of gas that a clump contains at a density higher than the gas mean density $\langle \rho_{gas} \rangle$ (hence with a \fft shorter than $\langle \tau_{ff} \rangle$), and we refer to this region of the clump as its inner region of mass $m_{inner}$.  
The gas mass fraction represented by the clump inner region, $m_{inner}/m_{gas}$, and the ratio between the mean \fft of the inner region  and the mean \fft of the clump as a whole, $\langle \tau_{ff, inner}\rangle/\langle \tff \rangle$, are shown in the bottom panel of Fig.~\ref{fig:denser}, for a range of $p$-values and $r_c/r_{clump}$ ratios identical to those used in the top panel.  
The inner region and the more diffuse outer region surrounding it contribute equally to the clump gas mass (i.e. $m_{inner}/m_{gas}=0.5$) when $p=1.5$.  For shallower density profiles, the inner region represents less than half of the clump mass (between 37\,\% and 50\,\%).  Combined to the weak contrast between the mean free-fall time of the clump gas and that of the inner region ($0.7 < \langle \tau_{ff, inner}\rangle/\langle \tff \rangle <1$), this explains why the inner region does not really "boost" the clump \sfr $SFR_{clump}$ when $p\leq 1.5$.  For instance, a shallow density gradient with $p=1$ increases the \sfr by 10\,\% at most (i.e. $\zeta \lesssim 1.1$; see Equation~\ref{eq:zeta}).  
In contrast, density profiles steeper than $p=1.5$ drive more than 50\,\% of the clump gas into the inner region in addition to yielding a greater contrast between the mean free-fall time of the clump gas and the mean free-fall time of the inner region.  This allows such clumps to achieve higher magnification factors $\zeta$ than those of shallow density profiles.  

\section{Dense-gas density gradient and dense-gas relation}\label{sec:dgrel}
\begin{figure}
\begin{center}
\epsscale{1.0}  \plotone{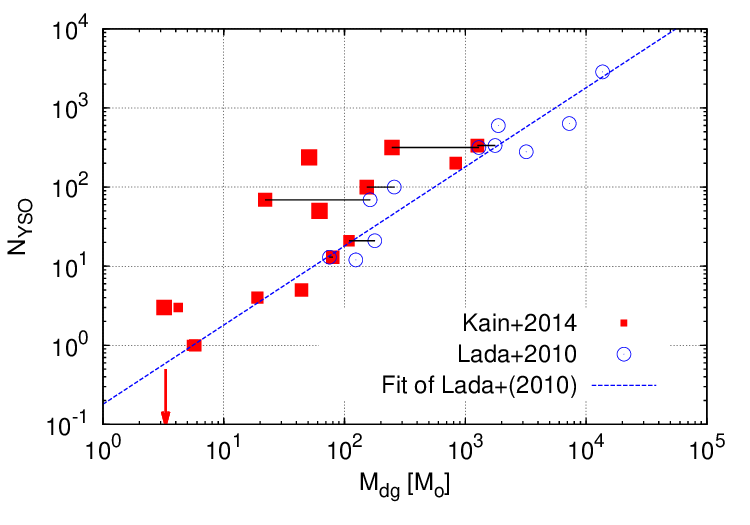}
\caption{Observed relation between the dense-gas content $M_{dg}$ of nearby molecular clouds and their YSO census $N_{YSO}$.  The red plain squares and the blue open circles depict the data of \citet{kai14} and \citet{lad10}, respectively.  Black horizontal lines connect the data points of clouds common to both studies: while their YSO numbers are identical, their respective dense-gas masses have been estimated based on different approaches (see text for details).  The vertical downward arrow indicates Chamaeleon~III, the one cloud in the sample of \citet{kai14} with  no YSO.  The dashed blue line represents the fit $N_{YSO}=0.18M_{dg}$ of \citet{lad10} to their data.  The size of the red squares is proportional to the steepness $p$ of the gas density profile, as estimated by \citet{kai14} from the slope of the $\rho$-PDF they have derived for their molecular clouds.}
\label{fig:dg}
\end{center} 
\end{figure}   

Figure \ref{fig:dg} shows as open circles the dense-gas relation inferred by \citet{lad10} for a sample of 11 molecular clouds with heliocentric distances of less than 500\,pc (see their Table 2).  $N_{YSO}$ and $M_{dg}$ are the cloud YSO number and cloud dense-gas mass, respectively.  The blue dashed line is the linear fit of \citet{lad10}: $N_{YSO} = 0.18 M_{dg}[\Ms]$.  The cloud dense-gas masses are extracted from a uniform set of infrared extinction maps, with the dense gas mass being defined as the cloud mass above a $K$-band extinction threshold of $A_K\simeq 0.8$\,mag.  This is equivalent to a gas surface density higher than $\Sigma_{gas} \simeq 116\,\Mspp$.  We have completed Figure~\ref{fig:dg} with the data set from \citet[][their Table~S1]{kai14}, which covers 16 molecular clouds closer than 260\,pc.  For the six clouds in common between both studies (Ophiuchus, Taurus, Lupus~I, Lupus~III, Corona-Australis, and the Pipe), the reported YSO numbers are identical and the corresponding data points in Figure~\ref{fig:dg} are connected by solid horizontal black lines.     Both studies, however, differ regarding how they estimate the cloud dense-gas content.   In \citet{kai14}, the dense-gas mass is based on a volume density threshold, rather than on a surface density threshold.  To do so, they devise a technique to convert the column density map of a cloud into an ensemble of hierarchical prolate spheroids, thereby approximating a three-dimensional map of the cloud hierarchical structure.  The volume and mass of the prolate spheroids are calculated, yielding the probability distribution function of the cloud gas volume density ($\rho$-PDF)\footnote{More specifically, in the case of \citet{kai14}, this is the probability distribution for the logarithm base $e$ of the gas volume density normalized by the cloud mean density (see below).}.  The cloud $\rho$-PDF inferred by \citet{kai14} is referred to as the "derived $\rho$-PDF" (to make it distinct from the true underlying $\rho$-PDF).  It is depicted as the black symbols in their Figures S1-S3.  Note that the derived $\rho$-PDF does not necessarily cover the full span in gas volume density of the cloud \citep[due to e.g. the limited range of column densities of the observational data; see][]{kai14}.

The $\rho$-PDF of a gas cloud is predicted to be either a log-normal, for supersonically turbulent, isothermal, non-self-gravitating gas \citep[e.g.][]{vaz94,pas98,sca98}, or a power-law, when self-gravity dominates \citep[this is equivalent to the gas developing a power-law density profile; see e.g.][]{sca98,kri11,gir14,koe19}.  The observed $\rho$-PDF of molecular clouds is often reported to be a combination of both \citep[see e.g.][]{kai09,schn15,koe19}.  \citet{kai14} fit both types of function to their derived $\rho$-PDF (solid and dashed lines, respectively, in their Figures~S1-S3).  To infer a three-dimensional estimate of the cloud dense-gas mass, they build on the fitted log-normal $\rho$-PDF to integrate the mass of all volume elements from a given threshold $s_{th}$ to infinity.  Here $s$ is the logarithmic mean-normalized density $s = ln(\rho/\rho_0)$, with $\rho_0$ the cloud gas mean density.  \citet{kai14} define the dense-gas as gas denser than $s_{th} = ln(\rho_{th}/\rho_0)=4.2$.  The average of the mean number densities of all clouds analysed by \citet{kai14} is  $n_0=100\,cm^{-3}$, corresponding to an averaged threshold number density $n_{th}=n_0\cdot e^{s_{th}}=6700\,\cc$ for the whole cloud sample.  The volume- and number-density thresholds ($\rho_{th}$ and $n_{th}$) present cloud-to-cloud variations, reflecting variations in the cloud mean density: $n_{th}$ ranges from $3.5\cdot10^3\,\cc$ (Corona Australis, $n_0=52\,\cc$) to $1.6 \cdot10^4\,\cc$ (Cham~II, $n_0=242\,\cc$) \citep[see Table S1 in][]{kai14}.  In the next section, we will show that these variations have little impact on the comparison between our model and the data.  By building on a three-dimensional density threshold, \citet{kai14} ignore the outer gas envelopes surrounding dense-gas clumps, which likely explains why the dense-gas masses of \citet{kai14} are on the average smaller than those of \cite{lad10}.    

The dense-gas relations of \citet{lad10} and \citet{kai14} also differ with respect to their correlation coefficient: linear fits to both data-sets provide $N_{YSO}=0.27\cdot M_{dg}^{0.94}$ (correlation coefficient $r=0.95$) and $N_{YSO}=0.55\cdot M_{dg}^{0.96}$ (correlation coefficient $r=0.84$), respectively.  That is, the sample of \citet{kai14} is characterized by a greater scatter than the sample of \citet{lad10} (see Figure~\ref{fig:dg}).  As we shall see through this section, the greater scatter in the data of \citet{kai14} is partly related to the gas density gradient inside the clouds and their dense-gas regions.

\begin{figure}
\begin{center}
\epsscale{1.0}  \plotone{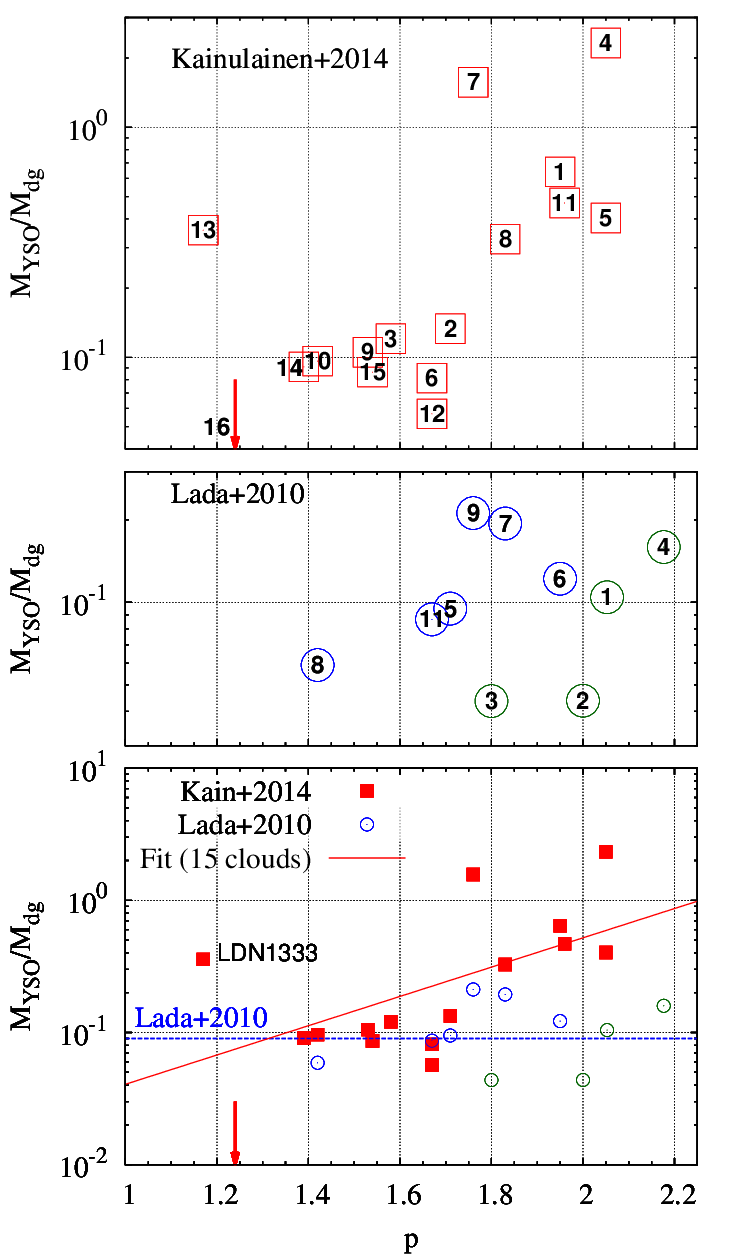}
\caption{Ratio $M_{YSO}/M_{dg}$ between the total mass in YSOs and the dense-gas mass of nearby molecular clouds, in dependence of the steepness $p$ of an equivalent power-law gas density profile $\rho_{gas}(r) \propto r^{-p}$.  $p$ stands for the gas density regime over which the cloud $\rho$-PDF can be approximated by a power-law.  Top panel: Sample of \citet{kai14} with the number-tagging corresponding to the line numbering in their Table~S1.  Middle panel: Sample of \citet{lad10} with the number-tagging corresponding to the line numbering in their Table~2.  Blue and green circles indicate different sources for the value of $p$ (see text for details).  Bottom panel: combination of both data sets along with the dense-gas relation of \citet{lad10} (blue horizontal dashed line).  The red solid line is a least-squares fit to the 15 clouds of \citet{kai14} (Cham~III is ignored considering its null YSO-census).  Note that the $y$-range extent differs from one panel to the other.  }
\label{fig:dgratio}
\end{center} 
\end{figure}   

When the $\rho$-PDF obeys a power-law, an equivalent power-law density profile $\rho_{gas}(r) \propto r^{-p}$ of the gas can be inferred.  Assuming a certain geometry for the gas reservoir (e.g. spherical or cylindrical), the steepness $p$ of the corresponding gas density profile is directly related to the slope of the $\rho$-PDF \citep{kri11,kai14}, or to that of the projected-density-PDF \citep{kri11,schn15}.  Using the power-law fit to their derived $\rho$-PDF, \citet{kai14} obtain     
the steepness $p$ of an equivalent spherically-symmetric gas density profile.  It is given as the $\kappa$ parameter in their Table~S1 (i.e. our $p\equiv$ their $\kappa$).  In what follows, we assume that the inferred steepness also stands for the dense-gas, even when the derived $\rho$-PDF does not probe significantly into it (e.g. Lupus~III and Cham~II, for which the derived $\rho$-PDF is limited to $s\lesssim s_{th}=4.2$: see their Figures~S1 and S2).  In Figure~\ref{fig:dg}, the size of the red squares depicting the data of \citet{kai14} is proportional to $p$, that is, larger symbols correspond to steeper gas density profiles inside molecular cloud structures.  \cite{par19} suggests that the scatter of the dense-gas relation of \citet{kai14} is not purely random, but is instead partly driven by $p$-variations.  She notes that, for a given dense-gas mass, clouds hosting a higher number of YSOs tend to have a higher $p$-value, although the effect is significant at the $1-\sigma$ level only (see her Figure~1).  Yet, the bottom panel of our Figure~\ref{fig:dgratio} strengthens her interpretation.  It shows the ratio between the YSO total mass $M_{YSO}$ and the dense-gas mass $M_{dg}$ of the clouds as a function of the steepness $p$ of the cloud density profile.   The total mass in YSOs is obtained under the assumption of a mean stellar mass $\langle m \rangle =0.5\,\Ms$, that is, $M_{YSO}[\Ms]=0.5 N_{YSO}$.  While the red plain squares depict the data of \citet{kai14}, the blue open circles depict the data of \citet{lad10} for which \cite{kai14} provide a $p$-value (i.e. the 6 clouds in common between both studies).  For 4 additional clouds (Orion A, Orion B, California and Perseus), we have estimated the steepness $p$ from the projected-density-PDFs ($\Sigma$-PDF) of \citet{lom15}.  Their Table~1 provides the index $n$ of the power-law $\Sigma$-PDF 
which is related to the steepness $p$ of an equivalent spherically-symmetric volume density profile as $p=1+2/n$ \citep{kri11}.  These four clouds are marked as green open circles.  To ease the identification of the clouds, the top and middle panels display the samples of \citet{kai14} and \citet{lad10}, respectively, with each point tagged with the line numbering in Table~S1 of \citet{kai14} (e.g.  Ophiuchus $\equiv 1$ and Cham~III $\equiv 16$) and Table~2 in \citet{lad10} (e.g. Orion~A $\equiv 1$ and Lupus~I $\equiv 11$). 

The horizontal dashed line in the bottom panel corresponds to the linear fit of \citet{lad10} shown in Figure~\ref{fig:dg} (i.e. $N_{YSO}/M_{dg}=0.18\,\Ms^{-1}\equiv M_{YSO}/M_{dg} = 0.09$).  The combined data of \citet{lad10} and \citet{lom15} do not show any particular trend: the ratio $M_{YSO}/M_{dg}$ remains about constant as $p$ increases.  The data of \citet{kai14}, however, present a different picture with the ratio $M_{YSO}/M_{dg}$ increasing with $p$.  A linear fit to their data gives $M_{YSO}/M_{dg}=0.003\cdot10^{1.1p}$ (solid red line; correlation coefficient $r=0.57$).  This shows that to increase its YSO census, a cloud can either increase its dense-gas mass, {\it or} steepen its gas density profile.  

Observationally, it is well-established that molecular clouds with a greater dense-gas content are more efficient at forming stars \citep{kai11,russ13}, hence the near-constancy of the $M_{YSO}/M_{dg}$ ratio \citep{lad10,lad12}.  This greater dense-gas content (e.g. more clumps and/or more filaments) is revealed by a shallower gas-density-PDF \citep{kai11,russ13,kai14}.  But what the data of \citet{kai14} in Figure~\ref{fig:dgratio} show is that such a pattern may also prevail {\it inside} dense-gas clumps.  That is, exactly as molecular clouds increase their \stf activity when a greater fraction of their gas is in the dense-gas regime, dense-gas clumps which locate a greater fraction of their gas in their high-density inner regions also raise their \stf activity.  Such clumps will present shallower $\rho$-PDFs than their less active siblings.  That clumps with steeper density profiles, hence shallower $\rho$-PDFs, yield more efficient \stf is as expected from semi-analytical computations as well as hydrodynamical simulations \citep{tan06,gir11a,cho11,elm11,fed13,par14p,par19}.  Two caveats, however, need to be kept in mind.  Firstly, when we consider the variations (or the absence thereof) of the ratio $M_{YSO}/M_{dg}$ as a probe into star formation, we implicitly assume that YSOs have formed exclusively inside the cloud dense gas.  Secondly, we have assumed that the steepness $p$ inferred by \citet{kai14} from the derived $\rho$-PDF also applies to the dense molecular gas, even when the derived $\rho$-PDF hardly covers the dense-gas regime (that is, when the derived $\rho$-PDF is limited to $s \simeq s_{th} = 4.2$).    

While the bottom panel of Figure~\ref{fig:dgratio} suggests that both the dense-gas content and the gas density gradient of molecular clouds are key drivers of their ability to form stars, we argue that the linear fit shown as the solid red line does not capture the underlying physics.  To quantify the evolution of the $M_{YSO}/M_{dg}$ ratio as a function of the steepness $p$, we need to rely on the magnification factor introduced by \cite{par19}.

Under the assumption that \stf takes place in dense gas only, we use Equation~\ref{eq:sfrclump2} to express the \sfr of the dense gas as a function of its mass, $M_{dg}$, and of its mean free-fall time $\langle \tau_{ff, dg} \rangle$:
\begin{equation}
SFR_{dg}=\zeta \effint \frac{M_{dg}}{\langle \tau_{ff,dg} \rangle}\;.
\label{eq:sfrdg}
\end{equation}

\begin{figure}
\begin{center}
\epsscale{1.0}  \plotone{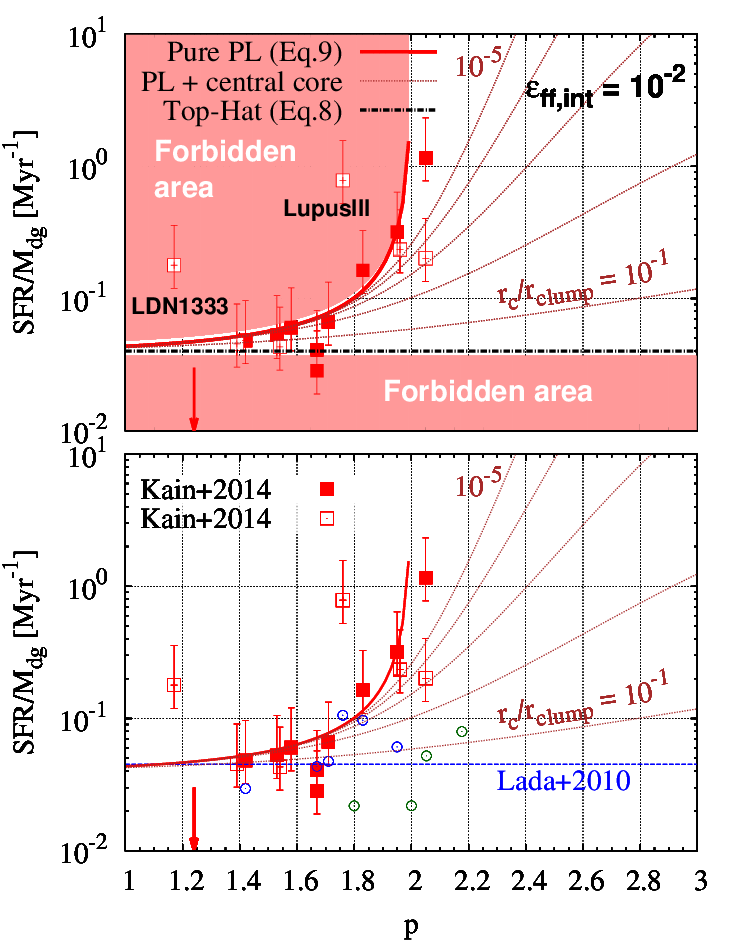}
\caption{Top panel: Ratio of the \sfr to the dense-gas mass $SFR/M_{dg}$ as a function of the steepness $p$ of power-law gas density profiles.  The lines correspond to the $\zeta$ factor presented in Fig.~\ref{fig:denser} (same color-coding), adopting $\effint=10^{-2}$ and $\langle \tau_{ff, dg} \rangle = 0.25$\,Myr.  The corresponding forbidden areas are highlighted in red.  The (plain and open) red squares depict the sample of \citet{kai14}.  Bottom panel: same as top panel but completed with the sample of \citet{lad10} and  their dense-gas relation (same color-coding as in the bottom panel of Figure~\ref{fig:dgratio}).  Note that the latter coincides with the top-hat prediction when $\effint \simeq 10^{-2}$.  }
\label{fig:newratio}
\end{center} 
\end{figure}

With $\zeta$ a function of both the steepness $p$ of the gas density profile and of the relative extent $r_c/r_{clump}$ of its central core, i.e. $\zeta = \zeta(p, r_c/r_{clump})$ (see top panel of Fig.~\ref{fig:denser}), it is clear that the ratio between the \sfr and the mass of the dense-gas cannot be constant, even for a given intrinsic \sfe per \fft and a given \fft of the dense gas.  The ratio $\effint / \langle \tau_{ff,dg} \rangle$ defines a lower limit corresponding to the case of a top-hat profile ($p = 0$) for which $\zeta = 1$ (see Equation~\ref{eq:zeta}):
\begin{equation}
\left(\frac{SFR_{dg}}{M_{dg}}\right)_{min}=\frac{\effint}{\langle \tau_{ff,dg} \rangle}\;.
\label{eq:ratioTH}
\end{equation}
With an estimate of $\langle \tau_{ff,dg} \rangle$, the comparison of Eq.~\ref{eq:ratioTH} with observational data for shallow density profiles will provide an estimate of $\effint$ (see below).
Any significant density gradient will raise the dense-gas ratio $SFR_{dg}/M_{dg}$ to a higher value, depending on $p$ and $r_c/r_{clump}$.   
When $p<2$, an upper limit is given by the case of a pure power-law gas density profile.
Combining Equations~\ref{eq:zeta} and \ref{eq:sfrdg}, we obtain:
\begin{equation}
\left(\frac{SFR_{dg}}{M_{dg}}\right)_{max,\,p<2}=\zeta \frac{\effint}{\langle \tau_{ff,dg} \rangle}= \frac{(3-p)^{3/2}}{2.6(2-p)} \frac{\effint}{\langle \tau_{ff,dg} \rangle}\;.
\label{eq:newratio}
\end{equation}
When $p>2$, how high the $SFR_{dg}/M_{dg}$ ratio becomes depends sensitively on the relative extent of the central core of the dense-gas clump, thereby reflecting the corresponding variations of the magnification factor $\zeta$ (see top panel of Fig.~1).  The top panel of Fig.~\ref{fig:newratio} illustrates the corresponding dense-gas ratios: from the case of a pure power-law to the case of a top-hat profile, with the intermediate cases of a power-law with a central core of relative extent $r_c/r_{clump}$.  We use the same $r_c/r_{clump}$ ratios as in Fig.~\ref{fig:denser}, and we set the \sfe per \fft  to $\effint=0.01$ and the dense-gas \fft to $\langle \tau_{ff,dg} \rangle=0.25$\,Myr (see below).  The corresponding forbidden areas (i.e. below the limit for a top-hat profile and above the limit for a pure power-law) are shaded in red.

We still need to estimate $\langle \tau_{ff,dg} \rangle$.    
We have seen above that, on the average, the number density threshold adopted by \citet{kai14} is $n_{th} \simeq 6700 \,\cc$ (that is, this is the averaged lower limit to the dense-gas number density).  The mean density of a centrally-concentrated clump is $3/(3-p)$ times the density $n_{th}$ at its edge \citep[Equation~7 in][]{par11c}.  With $p=1.67$ \citep[the median value of $p$ in the sample of ][]{kai14}, we thus get a clump mean number density $\langle n_{clump} \rangle = 2.3 \cdot 6700\,\cc = 1.5 \cdot 10^4\,\cc$ and a mean \fft  $\langle \tau_{ff,dg} \rangle=0.25$\,Myr.    

To compare the model predictions to the sample of \citet{kai14}, we convert the stellar masses $M_{YSO}$ used in Figure~\ref{fig:dgratio} into \stf rates averaged over a given star-formation time-span $t_{SF}$: 
\begin{equation}
SFR_{obs} = \frac{M_{YSO}}{t_{SF}} = \frac{\langle m \rangle N_{YSO}}{t_{SF}}\;.
\label{eq:sfrobs}
\end{equation}

In Figure~\ref{fig:newratio}, the red squares depict the data of \citet{kai14} assuming $\langle m \rangle = 0.5\,\Ms$ and a \stf duration $t_{SF}=2$\,Myr, which is the estimated lifetime of Class~II objects \citep{eva09}.  The distinction between plain and open symbols will be explained in the next section.  The error bars correspond to the upper and lower limits on the duration of the Class~II phase, that is, $t_{SF} \simeq t_{Cl~II} = 2 \pm 1$\,Myr \citep{eva09}.  If one excludes LDN1333, the data suggest a double-index power-law for the dense-gas relation, namely, a regime of near constancy as long as $p<1.7$, followed by a sharp rise.  
This slow-then-fast rise with $p$ shown by the data of \citet{kai14} is better captured by the permitted region defined in the top panel of Fig.~\ref{fig:newratio} than by a purely linear relation.  The permitted region actually accounts for both the near constancy of the data when $p<1.7$ and their rise when $p>1.7$.  At this stage, we can estimate the intrinsic \sfe per \fft by matching the top-hat prediction (black horizontal dash-dotted line) to the data for shallow density profiles (focusing on the plain squares; see Section~\ref{sec:kai}).  We adopt $\effint = 10^{-2}$. 
   
In the bottom panel, the horizontal line depicts the fit of \citet{lad10} $SFR/M_{dg} = 0.045\,\rm{Myr}^{-1}$ (with $t_{SF}=2$\,Myr).  It coincides with Equation~\ref{eq:ratioTH} (top-hat model) when $\effint \simeq 0.01$.

\section{Caveats about data and model} \label{sec:kai}

The great interest of the data of \citet{kai14} is that their dense-gas mass estimate builds on a {\it volume} density threshold.  Nevertheless, there are a few caveats to keep in mind.  

For some of the clouds, the dense-gas regime is not or little covered by the derived $\rho$-PDF.  This is the case of e.g. Chamaeleon~II and Lupus~III \citep[see Figures~S1 and S2, respectively, in][]{kai14}.  For such clouds, direct information about their $\rho$-PDF in the dense-gas regime is missing. 
A first caveat is thus that the dense-gas mass estimate sometimes stems from an extrapolation of the derived $\rho$-PDF {\it under the assumption of a Gaussian fit}, even though the actual behavior of the $\rho$-PDF in the dense-gas regime remains unknown.  Should a different function be fitted to the derived $\rho$-PDF, the dense-gas mass estimate may turn out to be different.  

A second caveat is that the slope of the $\rho$-PDF of these clouds is, over the dense-gas regime, uncertain, and so is the steepness $p$ of the equivalent dense-gas density profiles.  The $\rho$-PDF slopes given by \citet{kai14} are in fact measured over the density span for which the cloud is "structured" (that is, the $\rho$-PDF can be described by a power-law).  As this density span includes gas with a density $s<s_{th}=4.2$, an estimate of the $\rho$-PDF slope is available regardless of whether or not the dense-gas regime is well-sampled by the derived $\rho$-PDF.  The question here is therefore whether a $\rho$-PDF slope inferred at intermediate densities only, also applies to the dense-gas.  This is not necessarily the case.  For instance,  \citet{schn15} show that, in the molecular clouds MonR2 and NGC~6334, the projected-density-PDF becomes shallower when $n \gtrsim 10^4\,\cc$, thereby highlighting steeper radial density profiles (i.e. higher $p$) in the dense-gas regions of these molecular clouds.  The total $\rho$-PDF of the 16 clouds studied by \citet{kai14} actually shows a slight hint of becoming shallower in the dense-gas regime (see their Figure~S4).  Given that the $p$-values used in Figure~\ref{fig:newratio} are derived from the slope of the $\rho$-PDF, any uncertainty affecting the latter also affects the estimate of $p$.   

\begin{figure}
\begin{center}
\epsscale{1.0}  \plotone{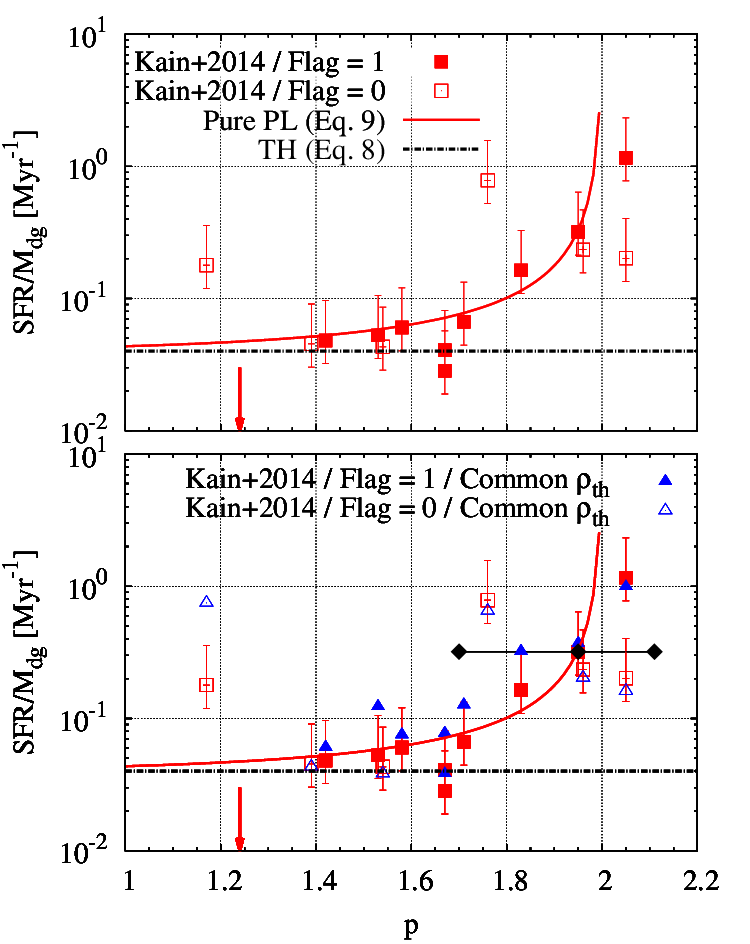}
\caption{Discussion of the uncertainties affecting the data used in Section~\ref{sec:dgrel}.  Top panel: solid red line, black dashed line and red symbols as in top panel of Figure~\ref{fig:newratio}.  The clouds in the sample of \citet{kai14} for which the dense-gas regime is poorly or not at all covered by their derived $\rho$-PDF are marked with open symbols ("Flag=0" in the key).  Bottom panel: same as in top panel, but completed with blue symbols which illustrate by how much the data points get vertically shifted if the dense-gas mass is estimated based on a volume density threshold $\rho_{th}$ common to all clouds (spherically-symmetric clumps are assumed; see text for detail).  The three black diamonds illustrate 3 different values of $p$ found in the literature for the Ophiuchus molecular cloud, from left to right: \citet{kau11}, \citet{kai14} and \citet{lom15}. }
\label{fig:wflag}
\end{center} 
\end{figure}

In Figures~\ref{fig:newratio} and \ref{fig:wflag}, we have identified the clouds whose dense-gas regime is poorly sampled by the derived $\rho$-PDF.  Specifically, we mark with open symbols ("Flag=0" in the key of Fig.~\ref{fig:wflag}) the clouds whose $\rho$-PDF shows at most one point with $s>s_{th}=4.2$.  For these clouds -- Cham~II (ID5), Lupus~III (ID7), LDN~134 (ID11), LDN1333 (ID13), LDN1719 (ID14), Musca (ID15), Cham~III (ID16), the steepness $p$ and the mass $M_{dg}$ of the dense-gas may not be well-constrained.  This is the case of the two clouds located in the upper forbidden area (LDN1333 (ID13) and Lupus~III (ID7)).  Their dense-gas content $M_{dg}$ may have been underestimated, as a result of $M_{dg}$ being  extracted from a Gaussian fit to the overall data, rather than based on a power-law fit.  
Alternatively, their location in a forbidden area may indicate different model parameters (see below).

A third caveat is that the threshold adopted by \citet{kai14}, $s_{th}=4.2$, applies to the dimensionless density $s = ln (\rho/\rho_0)$, with $\rho_0$ the cloud mean density.  Because not all clouds have the same mean density $\rho_0$, they are assigned different physical density thresholds $\rho_{th}=\rho_0 e^{s_{th}}$.  For instance, a higher-than-average mean density $\rho_0$ yields a higher-than-average density threshold $\rho_{th}$ which, in turn, lowers the dense-gas mass estimate.  The variations are not wide, however: the most diffuse and densest clouds in table~S1 of \citet{kai14} have mean number densities of, respectively, $n_0=52\,\cc$ (Corona Australis), and $n_0=242\,\cc$ (Chamaeleon~II).  We have estimated  by how much the points in the top panel of Figure~\ref{fig:wflag} get vertically shifted when the density threshold, instead of obeying $\rho_{th}=\rho_0 e^{s_{th}}$, is assigned the same value for all clouds, namely, $\rho_{th} = 700\,\Msppp \equiv n_{th}=10^4\,\cc$.  To do so, we use Equation~4 in \citet{par11c} which shows that, for a spherical clump with a pure power-law density profile, the mass of gas denser than a given volume density $\rho_{th}$ obeys $m_{dg} \propto \rho_{th}^{-(3-p)/p}$.  We can therefore write:
\begin{equation}
m_{dg,2} = m_{dg,1} \left( \frac{\rho_{th,2}}{\rho_{th,1}}\right)^{-\frac{3-p}{p}} 
\end{equation}
with $m_{dg,1}$ and $\rho_{th,1}$ the dense-gas mass and density threshold of \citet{kai14}, while $\rho_{th,2}$ is our adopted fixed density threshold ($\rho_{th} = 700\,\Msppp$) and $m_{dg,2}$ the corresponding dense-gas mass.  
The bottom panel of Figure~\ref{fig:wflag} compares our results (blue triangles) with the content of the top panel.  In all but two cases, the shift is not more significant than the uncertainties introduced by the duration of the YSO Class~II phase.  
          
Last but not least, uncertainties affecting the measured value of $p$ may be quite large as different methods may yield significantly different values.  We illustrate this point with the Ophiuchus molecular cloud for which \citet{kai14} quote $p=1.95$ in their table S1 (ID1).  Based on a cumulative-mass--size mapping of the cloud, \citet{kau11} derive $p \simeq 1.7$, while the slope of the projected-density-PDF of \citet[][$n=1.8$ in their table~1]{lom15} yields $p=2.1$ (p=2/n+1).  All three estimates are shown as the black diamonds in the bottom panel of Figure~\ref{fig:wflag}, showing the difficulty to ascertain whether Ophiuchus actually belongs to the permitted region.  

That data points are located in a forbidden area may also indicate the need for different parameters.  For instance LDN1333 and Lupus~III may require a higher \sfe per \fft (to raise the model towards the data points), or a longer \stf time-span $t_{SF}$ \citep[to lower the data points towards the model; e.g. ][]{bel13}.  An additional issue is that the cloud \sfr may have been varying as a function of time \citep[see e.g. Fig.~3 in][]{che20}.  One should keep in mind that the model predictions relate the \sfr and the gas density profile {\it at a given time}.  If the density profile and \sfr were different in the past, the measured \sfr (here averaged over the duration $t_{SF}$ of the \stf episode) then accounts for both the past and present \stf rates.  Yet, this is the present structure of the cloud which is mapped with the $p$-parameter.  If the \sfr has recently decreased, the measured ($t_{SF}$-averaged) \sfr is higher than the current one, thereby raising the data point in the parameter space.

We conclude that more data with lower uncertainties for the estimate of $p$ are needed, especially in the ascending part of Eq.\ref{eq:newratio}.  Estimates of the \sfr building on a short and recent time-span (i.e. quasi-instantaneous \stf rates) are also needed.

\section{Molecular Clouds of the Central Molecular Zone} \label{sec:cmz}
Molecular clouds of the Central Molecular Zone (hereafter CMZ, inner $\sim 200$\,pc of the Galaxy) have a mean $H_2$ density of about $10^4\,\cc$.  This is comparable to the density of the dense gas in clouds of the Solar neighborhood \citep[see][and references therein]{kau17a}.  CMZ clouds have been studied at an angular resolution of a few arc seconds (equivalent to $\simeq 0.1$\,pc spatial resolution) by \citet{kau17a}.  They find that \stf in some individual CMZ clouds is reduced by a factor of $10$ compared with the dense-gas relation established for nearby clouds by \citet{lad10} \citep[see Figure~8 in][]{kau17a}.  \citet{lu19} reach the same conclusion: with the exceptions of Sgr~C and Sgr~B2 which sits on the extrapolated dense-gas relation for nearby clouds, other CMZ clouds ($20\,km\,s^{-1}$, $50\,km\,s^{-1}$, G0.253+0.016 and Sgr~B1-off) have \stf rates about 10-times lower than the nearby-cloud prediction (see their Figure~6).  

In a follow-up study, \citet{kau17b} derive for each CMZ cloud a cumulative mass vs. bounding radius diagram, thereby yielding the slope of the underlying gas density profile (their Figure~3).    
In this Section, we take advantage of their study to revisit for CMZ clouds the relation between dense-gas mass, gas structure and \stf rate.  That is, we compare the available observational data to our new parameterization of the dense-gas relation, now defined as a permitted region enclosed in-between Eqs~\ref{eq:ratioTH} and \ref{eq:newratio} for a given intrinsic \sfe per free-fall time and a given \fft of the gas.    Based on Figure~3 in \citet{kau17b}, we adopt $p\simeq1.3$  for the clouds $20\,km\,s^{-1}$, $50\,km\,s^{-1}$, G0.253+0.016 (aka the Brick) and Sgr~B1-off, and $p\simeq2$ for Sgr~C and Sgr~B2\footnote{More precisely, we build on the fact that for a pure power-law density profile, the mass-size relation obeys $m(r)\propto r^{3-p}$, with m(r) the gas mass enclosed within the radius $r$, and has a logarithmic slope of $3-p$.  Based on figure~3 in \citet{kau17b}, we then infer $p=1.36$ for $20\,km\,s^{-1}$, $p=1.20$ for $50\,km\,s^{-1}$, $p=1.22$ for G0.253+0.016, $p=1.32$ for Sgr~B1-off, $p=2.08$ for Sgr~C and $p=2.00$ for Sgr~B2.}.  The cloud \stf rates and mass estimates (equivalent here to the dense-gas masses) are taken from \citet[][their table~7]{lu19}.  As discussed in \citet{par19} and in the previous section, this is the present gas density profile of a cloud or clump which drives its present \stf rate.  There may be little relation between the present structure of a cloud or clump and the \sfr it experienced, say, 1~Myr earlier.  This is because \stf modifies the gas density profile, especially in the cloud/clump inner regions where \stf is the fastest  \citep[see figures~3 and 4 in][]{par19}.  Since the gas density profile is modified, so is its impact on the \stf rate.  To relate the present structure of a cloud to its past \sfr obtained by averaging its total YSO mass over a time-span of 2\,Myr, as often done, including in our Figure~\ref{fig:newratio}, makes sense only if the \sfr has not varied significantly.  For CMZ clouds, \citet{lu19} alleviate this problem by focusing on tracers of recent, still deeply embedded, star formation, that is, ${\rm H_2O}$ masers and ultra-compact HII regions.  Older HII regions are discarded.  For both adopted tracers, they consider a \stf time-span of 0.3\,Myr, which is comparable to the \fft of gas with a number density of $10^4\,{\rm cm}^{-3}$.  This is also much shorter than the 2\,Myr time-span often assumed for nearby molecular clouds.  Their \sfr estimate therefore constitutes an almost instantaneous snapshot of the cloud \stf history, thereby allowing safer insights into the relation between cloud \sfr and gas density profile.  

\begin{figure}
\begin{center}
\epsscale{1.0}  \plotone{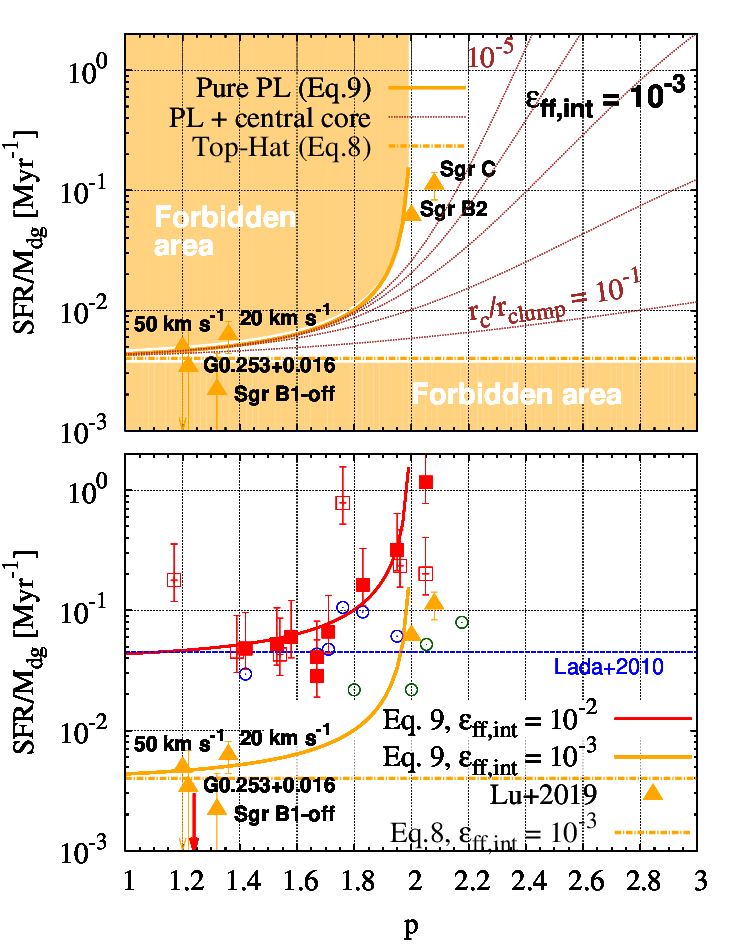}
\caption{Top panel: model lines as in top panel of Figure~\ref{fig:newratio} but for $\effint=10^{-3}$.  Corresponding forbidden areas are highlighted in orange.  The data for six clouds of the Central Molecular Zone are shown as orange plain triangles tagged with cloud names.    The \stf rates and dense-gas masses are from \citet{lu19}, and the steepnesses $p$ are from  \citet{kau17b}.  Bottom panel: bottom panel of Fig.~\ref{fig:newratio} completed with the CMZ cloud data. }
\label{fig:cmz}
\end{center} 
\end{figure}   

The bottom panel of Figure~\ref{fig:cmz} completes the samples of \citet{kai14} and \citet{lad10} with the above-described data for CMZ clouds.  Each orange plain triangle corresponds to one CMZ cloud, with the error bars depicting the \sfr uncertainties quoted by \citet{lu19} in their Table~7.  The solid orange line depicts the upper limit to the dense-gas relation adopted for nearby clouds shifted down by a factor of 10 \citep{kau17b,lu19}.  That is, this is Equation~\ref{eq:newratio} with $\effint=10^{-3}$.  The horizontal dash-dotted orange line corresponds to the case of a top-hat profile with $\effint=10^{-3}$.  The top panel illustrates the corresponding forbidden areas.  Within the uncertainties, CMZ clouds are located in the permitted area associated to $\effint = 10^{-3}$ and $\langle \tau_{ff,dg} \rangle=0.25$\,Myr. 
Per unit (dense) gas mass, Sgr~C and Sgr~B2 form stars at a pace about 10-times faster than the other CMZ clouds $20\,km\,s^{-1}$, $50\,km\,s^{-1}$, G0.253+0.016, and Sgr~B1-off.  Figure~\ref{fig:cmz} suggests that the more vigorous \stf activity in Sgr~C and Sgr~B2 stems -- at least in part -- from their steeper gas density profile.  

We can also compare CMZ and nearby clouds.  The bottom panel of Figure~\ref{fig:cmz} shows that the $SFR/M_{dg}$ ratio of the clouds $20\,km\,s^{-1}$, $50\,km\,s^{-1}$, G0.253+0.016, and Sgr~B1-off is 10-times lower than that of nearby molecular clouds {\it of similar density profile steepness} ($p \lesssim 1.6$).  Note that for shallow density profiles, the relative extent of the central core $r_c/r_{cloud}$ hardly modifies the $\zeta$ factor \citep[see top panels in Figure~5 of][and Figure~\ref{fig:denser} in this paper]{par19}.  The difference between the dense-gas relations for nearby and CMZ clouds cannot therefore be ascribed to CMZ clouds having a larger central core.  That the gas density profile must be taken into account when comparing the \stf activities of CMZ and \Son clouds was already emphasized by \citet{kau17b}.  However, Figure~\ref{fig:cmz} shows that the shallow density gradient of the least-active CMZ clouds is not enough to explain their low \stf activity.  
Figure~\ref{fig:cmz} rather suggests that either the intrinsic \sfe per \fft $\effint$ in CMZ clouds is 10-times lower than in nearby clouds, or that the $SFR_{dg}/M_{dg}$ ratio depends on a third, yet unknown, parameter in addition to the magnification factor $\zeta$ and the intrinsic \sfe per \fft $\effint$.  That is, the dense-gas relation should be re-written 
\begin{equation}
\frac{SFR_{dg}}{M_{dg}}= V_X \zeta \frac{\effint}{\langle \tau_{ff,dg} \rangle}\,,
\label{eq:vx}
\end{equation}
with $V_X$ a vertical shift of still unknown origin. \\
 
That star formation may not take place exclusively in the dense gas may also contribute to explaining the vertical shift between nearby and CMZ clouds.  Let us assume that in nearby clouds only 10 per cent of the \stf activity actually takes place in the dense gas.  That would decrease by a factor of 10 the \stf rates that we have adopted for the dense gas of nearby clouds, and both CMZ and nearby clouds would then be characterized by the same intrinsic \sfe per freefall time, i.e. $\effint=10^{-3}$.    
It is interesting to note that this value is also close to the "floor" of \stf efficiencies per free-fall time measured by \citet{lee16} for Galactic Giant Molecular Clouds.  That would also imply that the nearby clouds of \citet{kai14} form 90 per cent of their stars at densities lower than that of the dense gas.  In turn, this would require the definition of another gas density threshold for star formation.  Recently, \citet{bur19} have proposed that \stf takes place over the gas density range covered by the power-law part of the $\rho$-PDF. In their proposed model, the transition density from the lognormal regime to the power-law one depends on the interplay between gravity, turbulence and stellar feedback \citep[see also][]{aud19}. 
 
\begin{figure}
\begin{center}
\epsscale{1.0}  \plotone{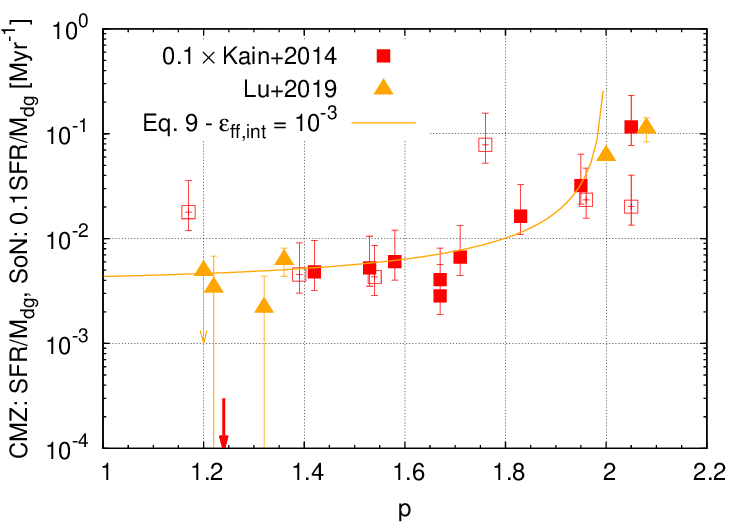}
\caption{Ratio between the \sfr and the dense-gas mass in dependence of $p$.  Orange symbols stand for the CMZ cloud data of \citet{lu19}.  Red symbols stand for the Solar-neighborhood (SoN) cloud data of \citet{kai14}, shifted down by a factor of 10.  Plain and open red squares have the same meaning as in the top panel of Figure~\ref{fig:wflag}.  The solid line shows Equation~\ref{eq:newratio} with $\effint=10^{-3}$. }
\label{fig:cmzson}
\end{center} 
\end{figure}

The bottom panel of Figure~\ref{fig:cmz} also suggests that the location of Sgr~C and Sgr~B2, close to the dense-gas relation of \citet{lad10} for nearby clouds, should be interpreted with caution: it may not imply that Sgr~C and Sgr~B2 are more akin to nearby clouds than to their CMZ siblings with shallower density profiles.  If we compare the data of \citet{kai14} and \citet{lu19}, Figure~\ref{fig:cmz} shows that around $p=2$, the $SFR/M_{dg}$ ratio of nearby clouds is higher than that of Sgr~C and Sgr~B2.  This suggests that the vertical shift between nearby and CMZ clouds noticed for $p \lesssim 1.5$  also stands for steeper gas density profiles.  To ascertain this, Figure~\ref{fig:cmzson} presents the data of \citet{lu19} and \citet{kai14} together, but with the data for nearby clouds being shifted down by a factor of 10.  Both data sets now remarkably occupy the same narrow region of the ($p$, $SFR/M_{dg}$) parameter space.  That is, once the variations of $SFR/M_{dg}$ with $p$ are taken into account, the dense-gas relation of CMZ clouds remains distinct from that of nearby clouds, including for Sgr~C and Sgr~B2.  The difference between both relations amounts to a factor of $10$, as quoted in previous studies \citep{kau17b,lu19}.  It may be due to either a ten-times lower $\effint$,  a yet unknown parameter (i.e. $V_X=0.1$ in Equation~\ref{eq:vx}), or nearby clouds forming only 10 per cent of their stars in dense gas.  For each cloud sample, the steepness $p$ of the dense-gas density profile is the key parameter yielding a range of $SFR/M_{dg}$ ratios.  \\
  
\section{Summary and Conclusions}\label{sec:conc}
We have presented a new parameterization of the relation between mass and \sfr of dense-gas clumps, taking into account the impact of their gas density gradient.  Specifically, we have combined the dense-gas relation with the magnification factor $\zeta$ introduced by \citet{par19}.  This factor quantifies by how much the gas density gradient of a centrally-concentrated clump enhances its \sfr compared to the \sfr that the clump would experience should it be of uniform density (i.e. with a top-hat profile; see Equation~\ref{eq:zetaratio}).  For shallow density profiles, the clump \sfr is barely higher than for its top-hat equivalent and $\zeta \gtrsim 1$.  Steeper density profiles, however, can yield $\zeta >> 1$ due to the greater mass fraction of gas in the clump inner regions where the gas \fft is shorter than for the clump as a whole.  

We consider spherically-symmetric dense-gas clumps.  
For a pure power-law density profile $\rho_{gas}(r) \propto r^{-p}$, the magnification factor $\zeta$ can be defined for $p<2$ and obeys Equation~6 in \citet{par19} (quoted as Equation~\ref{eq:zeta} in this paper).  To define the magnification factor for any $p$-value, a central core of finite density must be added to the clump gas density profile.  Compared to the case of a pure power law, a central core smooths the gas density contrast between clump center and clump edge, thereby decreasing the clump magnification factor \citep[see top panel of Figure~\ref{fig:denser} where shaded areas represent the corresponding forbidden regions, either because $\zeta$ is higher than for a pure power-law gas density profile (Eq.~\ref{eq:zeta}), or because $\zeta < 1 $, i.e. lower than for a top-hat profile; see also][]{par19}. 

The magnification factor and the dense-gas ratio "take-off" around $p=1.5$.  This is because for density profiles steeper than $p=1.5$, more than half of the gas mass has a density higher than the gas density averaged over the whole clump (see bottom panel of Figure~\ref{fig:denser}).  That is, for $p>1.5$, more than half of the clump gas mass forms stars at a faster pace/with a shorter free-fall time than expected based on the clump gas mean density. \\   

We have applied the concepts of intrinsic \sfe per \fft $\effint$ and magnification factor $\zeta$ introduced by \citet{par19} to redefine the relation between \sfr and mass of the dense gas (Equation~\ref{eq:sfrdg}).  This way of doing allows one to disentangle the respective contributions of the \sfe per \fft and of the gas density gradient to the \stf rate.  Through the $\zeta$ factor, the steepness $p$ of the gas density profile and the relative extent $r_c/r_{clump}$ of its central core constitute the newly-introduced parameters in the dense-gas relation.    

A top-hat density profile yields a natural lower limit to $SFR_{dg}/M_{dg}$, i.e. $SFR_{dg}/M_{dg} = \effint/\langle \tau_{ff,dg} \rangle$ (Eq.~\ref{eq:ratioTH}).  A pure power law yields an upper limit on $SFR_{dg}/M_{dg}$ when $p<2$ (Eq.~\ref{eq:newratio}).  For a given intrinsic \sfe per \fft $\effint$ and dense-gas \fft $\langle \tau_{ff,dg} \rangle$, the forbidden regions in the $(p,\zeta)$ parameter space yield forbidden regions in the $(p,SFR_{dg}/M_{dg})$ parameter space.  These are highlighted in red in the top panels of Figs~\ref{fig:newratio} ($\effint=10^{-2}$ and $\langle \tau_{ff,dg} \rangle=0.25$\,Myr) and \ref{fig:cmz} ($\effint=10^{-3}$ and $\langle \tau_{ff,dg} \rangle=0.25$\,Myr).  

We have compared our model predictions to nearby cloud observational data.  We have used the data of \citet[][their table S1]{kai14}, as these build on a gas volume density threshold, and we have assumed that nearby clouds form the bulk of their YSOs in their dense gas (but see below).  We show that the dense-gas regions of the nearby clouds surveyed by \citet{kai14} present a higher $SFR/M_{dg}$ ratio when $p \simeq 2$ than when $p\simeq1.5$ (top panel of Figure~\ref{fig:dgratio}).  Most of the data of \citet{kai14} in fact belong to the permitted area defined by our model for $\effint =10^{-2}$ (see Fig.~\ref{fig:newratio}).

In \citet{kai14}, the slope $-p$ of the gas volume-density profile is inferred from the slope of the gas volume-density-PDF that they derive\footnote{The steepness $p$ that they infer is given as the $\kappa$ parameter of their table S1.}.  For some of the clouds, the derived $\rho$-PDF does not cover the dense-gas regime.  This uncertainty, however, does not affect our conclusions (top panel of Figure~\ref{fig:wflag}), and may explain the presence of LDN1333 and Lupus~III in the upper forbidden area.   Also, because the density threshold adopted by \citet{kai14} is a dimensionless one (it is defined as the logarithmic mean-normalized density $s_{th}=ln(\rho_{th}/\rho_0)=4.2$, where $\rho_0$ is the cloud mean density), the corresponding volume density threshold $\rho_{th}$ varies from cloud to cloud.  Again, we have checked that this has no impact on our conclusions (bottom panel of Figure~\ref{fig:wflag}).  

We have also compared the predictions of our model to the observational data available for the dense-gas clouds of the Central Molecular Zone (CMZ).  Combining the data of \citet{lu19} and \citet{kau17b}, we find that CMZ clouds too  present an increase of their $SFR/M_{dg}$ ratio with $p$.  Specifically, Sgr~C and Sgr~B2, the two CMZ clouds with the steepest gas density gradient ($p \simeq 2$), are also the CMZ clouds presenting the highest $SFR/M_{dg}$ ratio.  This one is about  $10$-times higher than for CMZ clouds whose density gradient is shallower, namely,  $20\,km\,s^{-1}$, $50\,km\,s^{-1}$, G0.253+0.016 (aka the Brick) and Sgr~B1-off (Figure~\ref{fig:cmz}).  Such an increase is compatible with model predictions for $\effint=10^{-3}$, although strongly centrally-peaked gas density profiles for Sgr~C and Sgr~B2 are required (see Figure~\ref{fig:cmz}).   
In any case, if an additional, yet to be unveiled, parameter contributes to differentiating the \stf activity of Sgr~C and Sgr~B2 from that of their less active CMZ counterparts, then the difference in \sfr that this extra parameter has to account for is smaller than a factor of 10, since the difference in density gradient already contributes a factor of a few. 

We have emphasized that the presence of Sgr~C and Sgr~B2 on the extension of the dense-gas relation established by \citet{lad10} for nearby clouds does not necessarily imply that they are more akin to nearby clouds than to other CMZ clouds.  
When the data of \citet{lu19} are compared with those of \citet{kai14}, Sgr~C and Sgr~B2 detach themselves from nearby clouds of similar steepness $p$ (Figure~\ref{fig:cmz}).    

We actually find  the $(p, SFR/M_{dg})$ relation of nearby clouds to be about 10-times higher than that of CMZ clouds (see Figure~\ref{fig:cmzson}, where the nearby-cloud data have been scaled down by a factor of 10).  Either the intrinsic \sfe per \fft $\effint$ of the dense CMZ clouds is 10 times lower than for the nearby cloud dense gas, or nearby clouds form 10 per cent only of their stars in dense gas, or a third parameter - on top of $\effint$ and $\zeta(p, r_c/r_{clump})$ - contributes to the dense-gas relation (Eq.~\ref{eq:vx}).  

It is highly desirable to accumulate data for other \sfing environments, that is, to go beyond the CMZ and beyond the  immediate \Son (recall that the molecular clouds of \citet{kai14} are closer than 260\,pc).  This calls for (1) dense-gas masses defined based on a volume-density criterion, (2) reliable steepnesses $p$ measured over the dense-gas regime, and (3) \stf rates measured over short and recent \stf time-spans since the observed gas spatial distribution is best related to its {\it current} \stf rate.  This last aspect is to be kept in mind when putting to the test \sfing regions hosting multiple stellar populations \citep[e.g. the Orion Nebula Cluster, for which three subsequent \stf episodes have been detected so far;][]{bec17}.  More data will be helpful not only to further test the impact of a gas density gradient as predicted by our model, but also to investigate whether the intrinsic \sfe per \fft $\effint$ varies from one environment to another.  For this, \sfing regions with shallow density profiles will be the best targets since the impact of their gas density gradient is negligible (i.e. $\zeta \simeq 1$).  \\



\acknowledgments
G.P. acknowledges funding by the Deutsche Forschungsgemeinschaft (DFG, German Research Foundation -- Project-ID 138713538 -- SFB 881 ("The Milky Way System", subproject B05).  We thank an anonymous referee for their constructive criticisms which have improved the paper.  This research has made use of NASA 's Astrophysics Data System.

\appendix
Equation~\ref{eq:zeta} shows that, in case of a pure power-law density profile, the magnification factor $\zeta$ cannot be defined for $p\geq2$.  This is because the corresponding density profiles yield an infinite \sfr at the clump center, thereby preventing one from defining the total \sfr of the clump and the corresponding magnification factor.  With $\rho_{gas}(r)\propto r^{-p}$, the \sfr at the clump center obeys
\begin{equation} 
\lim_{r \to 0} \effint \frac{dm_{gas}(r)}{\tau_{ff}(r)}  \propto \lim_{r \to 0} \frac{r^{3}\rho_{gas}(r)}{\sqrt{\frac{1}{\rho_{gas}(r)}}}   \propto \lim_{r \to 0} \frac{r^{3-p}}{r^{p/2}} 
\end{equation} 
with $dm_{gas}(r \to 0)$ the gas mass element at the clump center.  Note that the definition of a finite clump mass requires $p<3$.  As $r$ approaches $0$, both the enclosed gas mass  $dm_{gas}(r \to 0)$ and the local \fft $\tff(r \to 0)$ approach zero.  Depending on which function approaches zero the fastest, the \sfr at the clump center is either zero or infinity.  If $3>p>2$, the \fft approaches zero faster than the central gas mass element, which yields an infinite \sfr at the clump center and an infinite magnification factor $\zeta$.  In contrast, for shallower density profiles ($p<2$), the central gas mass element approaches zero faster than the freefall time and  the \sfr at the clump center is zero.  This allows one to define the clump total \sfr and the magnification factor, as shown in the top panel of Figure~\ref{fig:denser}.  In case of a power-law gas density profile with a central core, the \fft at the clump center differs from zero, yielding finite \stf rates for the clump center and the clump as a whole.  For such cases, the magnification factor can always be defined, regardless of the steepness $p$ of the gas density profile.







\end{document}